\documentstyle[12pt]{article}

\input{psfig}
\large
\newcommand{\be}{\begin{eqnarray}}
\newcommand{\ee}{\end{eqnarray}}
\newcommand\del{\partial}

\begin{document}
\setlength{\baselineskip}{21pt}
\pagestyle{empty}  
\vfill                                                                          
\eject                                                                          
\begin{flushright}                                                              
SUNY-NTG-96/41
\end{flushright}                                                                
                                                                                
\vskip 2.0cm 
\centerline{\Large  Yang-Lee zeros of a random matrix model for QCD at
finite density}
\vskip 2.0 cm                                                                   
\centerline{M.A. Halasz$^{\rm a}$,  A.D. Jackson$^{\rm b}$, 
and J.J.M. Verbaarschot$^{\rm a}$}
\vskip 0.2cm            
\centerline{$^{\rm a}$ Department of Physics, SUNY, Stony Brook, New York 11794}
\centerline{$^{\rm b}$ Niels Bohr Institute, Blegdamsvej 17, Copenhagen,
DK-2100, Denmark}
\vskip 2cm                                                                   
                                                                                
\centerline{\bf Abstract}
We study the Yang-Lee zeros of a random matrix partition function with the
global symmetries of the QCD partition function.  We consider both 
zeros in the complex chemical potential plane and in the 
complex mass plane.   In both cases we find that the zeros 
are located on a curve.  In the thermodynamic limit, the zeros 
appear to merge to form a cut.  The shape of this limiting curve 
can be obtained from a saddle-point analysis of the partition function. 
An explicit solution for the line of zeros in the complex chemical 
potential plane at zero mass is given in the form of a transcendental 
equation.
\vfill                                                                          
\noindent                                                                       
\begin{flushleft}
November 1996
\end{flushleft}
\eject
\pagestyle{plain}

\noindent

1. Since the work of Yang and Lee \cite{frank}, zeros of the partition function
have become an important tool for the study
of phase transitions.  (See, e.g., \cite{huang,itzykson,shrock}.) 
Most notably, Yang and Lee proved the theorem that
the thermodynamic limit of the free energy is analytic
in any region of the complex fugacity plane which contains no zeros. 
Below, we use the term Yang-Lee zeros as a generic name for zeros 
of the partition function. (Note, however, that zeros in the complex
temperature plane were first discussed in \cite{fisher,katsura}.)
Historically, Yang-Lee zeros have been studied primarily within 
the framework of statistical mechanics.  Yang-Lee zeros of the Ising 
model have received particular attention.  (See \cite{shrockising} 
for recent work on this topic.)  Recently, however, Yang-Lee zeros 
were studied in the context of the lattice QCD (QED) partition function 
for non-zero chemical potential 
\cite{vink,barbourqed,barbourbell,barbourmu,barbourmass}. 
Because of the difficulty of simulating the partition function of 
lattice QCD for non-zero chemical potential \cite{everybody}, these 
studies are still in an exploratory stage.  In this work we study 
Yang-Lee zeros for a much simpler random matrix partition function 
which possesses the global symmetries of the QCD partition function. 
This enables us to obtain the Yang-Lee zeros both numerically and 
analytically.
 
The continuum Euclidean QCD partition function for a quark mass matrix, 
$m$, and chemical potential $\mu$ can be written as
\be
Z(m, \mu) = \left < \det(\gamma\cdot D +m + i\mu \gamma_0) \right >_{S_{QCD}},
\label{QCDpart}
\ee
where $\gamma\cdot D$ is the Euclidean Dirac operator and $\gamma_\mu$ 
are the Euclidean Dirac matrices.  The average is over the Yang-Mills 
action.  In lattice QCD, the chemical potential is incorporated by 
including a factor $e^\mu$ for links forward in time
and a factor $e^{-\mu}$ for links backward in time.  Since the lattice 
QCD Dirac operator is a finite matrix, the partition function
is a polynomial in $m$ and $e^\mu$.  This makes it possible to study zeros 
of the partition function in the complex fugacity plane.  The difficulty 
with lattice studies is that the coefficients of this polynomial are 
necessarily obtained numerically.  This is problematic because the 
zeros of a high-order polynomial are notoriously  sensitive to the 
values of its coefficients.

As is well known, the zeros of $Z$ in $m$ are closely related to the chiral
order parameter \cite{vink}. If these zeros are located at $z_k$, the 
partition function (\ref{QCDpart}) can be written as
\be
Z(m) = \prod_k(m-z_k) \ .
\ee
Therefore, the chiral condensate (for one flavor) is given as 
\be
\Sigma(m) \equiv \frac 1N \del_m \log Z(m) = \frac 1N \sum_k \frac 1{m-z_k}.
\ee
We consider $\Sigma(m)$ as a function of complex $m$. 
Provided that these zeros merge to form a cut in the thermodynamic 
limit,  $\Sigma(m)$ will show a discontinuity each time $m$ crosses 
such a line.  Similarly, the `baryon density' defined as $n_B \equiv 
\del_\mu \log Z$ shows a discontinuity if $\mu$ crosses a line of zeros 
in the complex $\mu$-plane.

2. In this work we study the Yang-Lee zeros for a random matrix model of 
the QCD partition function. In essence, we replace the matrix elements of
the Dirac operator by Gaussian distributed random variables consistent
with the global symmetries of the QCD partition function (see \cite{verb}
for a review).  
For $N_f$ flavors this partition 
function is defined as
\be
Z(m,\mu) = \int D C P(C) \prod_f^{N_f}
\det \left ( \begin{array}{cc} m_f & iC + \mu\\
iC^\dagger +\mu & m_f \end{array} \right) \ ,
\label{zmu}
\ee
where $C$ is an arbitrary complex $n\times n$ matrix and $DC$ the Haar measure.
The probability distribution $P(C)$ is given by
\be
P(C) = \exp (-n {\Sigma^2} {\rm Tr} C C^\dagger) \ .
\ee
We emphasize that this partition 
function is a $schematic$ model of the lattice QCD partition function.  
For example, it does not reproduce the zero temperature Fermi-Dirac 
distribution.  However, this partition function shares certain important 
features with the QCD partition function.  We mention four properties: 
1. This model shows a chiral phase transition as a function
of the chemical potential.  Chiral symmetry is broken at zero chemical 
potential and is restored above some critical $\mu$.  At $\mu = 0$ the 
chiral condensate is $\Sigma$.
2. The Dirac operator is non-hermitean with eigenvalues distributed
in the complex plane.
3. As first shown by Stephanov \cite{Stephanov2} and confirmed in \cite{Janik}, 
the quenched limit of this partition function is
obtained as the limit $N_f\rightarrow 0$ of (\ref{zmu}) with
the fermion determinant replaced by its absolute value.
4. If $\mu = i \pi T$, it can be interpreted as a model of the QCD partition
function with only the lowest Matsubara frequency included 
\cite{JV,Tilo,sener}. 
This model shows a second order phase transition at $\pi T\Sigma = 1$. 

Obviously, the partition function (\ref{zmu}) is a polynomial in $m$ 
and $\mu$.  The advantage of studying Yang-Lee zeros in this model 
is that the coefficients of the polynomial can be obtained analytically. 
This eliminates a major source of error in the calculation.  Here, we 
restrict our attention to a single flavor, and we adopt units in which 
the parameter $\Sigma$ is equal to 1. 

Using standard methods \cite{JV,Tilo,Stephanov1}, this partition function 
can be rewritten as
\begin{eqnarray}
\label{epf}
Z_{N}(m,\mu) \sim  \int d\sigma \, d\sigma^*
e^{-n \sigma \sigma^* }
{\det}^{n} \left( \begin{array}{cc}  \sigma+m  & \mu  \\
    \mu    & \sigma^* + m
\end{array} \right ) \ . 
\end{eqnarray}
In the thermodynamic limit this partition function can be evaluated 
by a saddle point approximation.  We obtain a non-trivial saddle 
point 
\be
\sigma^* &=&\sigma,\nonumber \\
\sigma(m+\sigma)^2 -\mu^2 \sigma &=& m+ \sigma \ .
\label{cubic}
\ee
For $\mu>0.527\cdots$, the trivial saddle point, $\sigma=0$, is dominant 
for $m=0$ \cite{Stephanov2}.  Chiral symmetry is restored above this point. 
The point at which the phase change occurs is determined by the solution 
to a transcendental equation.  For complex $\mu$ 
it is given by
\be
{\rm Re}[ \mu^2 + \log(\mu^2) ] = -1 \ .
\label{trans}
\ee
We wish to note two special solutions of this equation, $\mu = 0.527\cdots$
discussed above and $ \mu = i$.  The latter solution corresponds to the 
critical temperature $\pi T = 1$ found in \cite{JV}.  We will show below 
that the zeros in $\mu$ of $Z(m=0,\mu)$ for finite $n$ lie along 
this curve.  The discriminant of the cubic equation (\ref{cubic}) is given 
by
\be
D_3 = \frac 1{27}( m^4 \mu^2 - m^2(2\mu^4 - 5\mu^2 -\frac14) +(1+\mu^2)^3) \ .
\label{discriminant}
\ee
We will also show that the endpoints of a line of Yang-Lee zeros are given
by the points where this discriminant vanishes.

The partition function (\ref{epf}) can be written as a polynomial in $\mu$
and $m$ by a straightforward expansion of the determinant in powers of 
$\mu$ and $m$.  With the help of a number of combinatorial identities, 
we arrive at
\begin{eqnarray}
\label{series_1}
Z(m,\mu)=\frac{\pi N!}{N^{N+1}} \sum\limits_{k=0}^N \sum\limits_{j=0}^{N-k}
\frac{(Nm^2)^k}{(k!)^2} \frac{(-N\mu^2)^j}{j!} \frac{(N-j)!}{(N-j-k)!} \ .
\end{eqnarray}
As a special case we mention $m = 0$, for which the partition function, 
\be
Z(0,\mu) = \frac{\pi N!}{N^{N+1}} \sum\limits_{j=0}^N 
\frac{(-N\mu^2)^j}{j!} \ ,
\ee
is simply a truncated exponential.

3. In this section we evaluate the Yang-Lee zeros of the partition function
(\ref{series_1}). First, we consider the partition function as a polynomial 
in $m$.  Fig.\ 1 shows results for $\mu = 0$, $\mu = 0.5$, and $\mu = 0.6$.
We have calculated the zeros for different values of $n$, e.g., $n=48$, 
$n=96$, and $n=192$.  The results for $n=192$ are represented by the points 
in the figure.  Of course, the exact location of the zeros is extremely 
sensitive to numerical round-off errors. Thus, the present results were 
obtained with the help of a multi-precision package \cite{bailey}.  
Typically, we performed our computations with 100-500 digits accuracy.

\begin{figure}[htb]
\psfig{figure=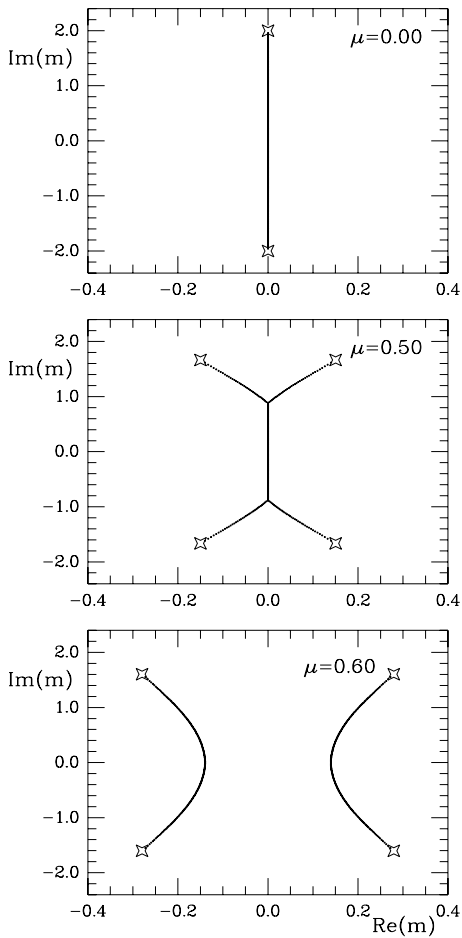,width=60mm}
\caption{The zeros of the partition function in the complex $m$ plane
for $\mu =0$ (upper), $\mu = 0.50$ (middle) and $\mu = 0.60$ (lower) 
for $n = 192$.  The zeros of the discriminant of the cubic equation 
(\ref{cubic}) are denoted by stars.}
\label{lym}
\vspace*{-0.5cm}
\end{figure}

The zeros fall on a curve and are regularly spaced\footnote{If the numerical
accuracy is not sufficient, one typically observes that the line of zeros ends
in a circle.}.  
From the saddle point analysis of the partition function it
is clear that the condition that the free energy of two different saddle
point solutions
coincides imposes a single  condition on the complex variable $m$.
This explains that, in
the thermodynamic limit, the zeros are located on a curve in the complex
plane (A similar argument has been given for the Ising model 
\cite{shrock}.).  If we increase the order of the polynomial by a factor of 
$2$, we find that half of the zeros are close to those of the lower-order 
polynomial.  The other zeros half are roughly mid-way between adjacent zeros 
of the lower-order polynomial.  This leads us to the conclusion that the 
zeros become dense and lead to a cut in the complex $m$ plane in the 
thermodynamic limit.

\begin{figure}[htb]
\psfig{figure=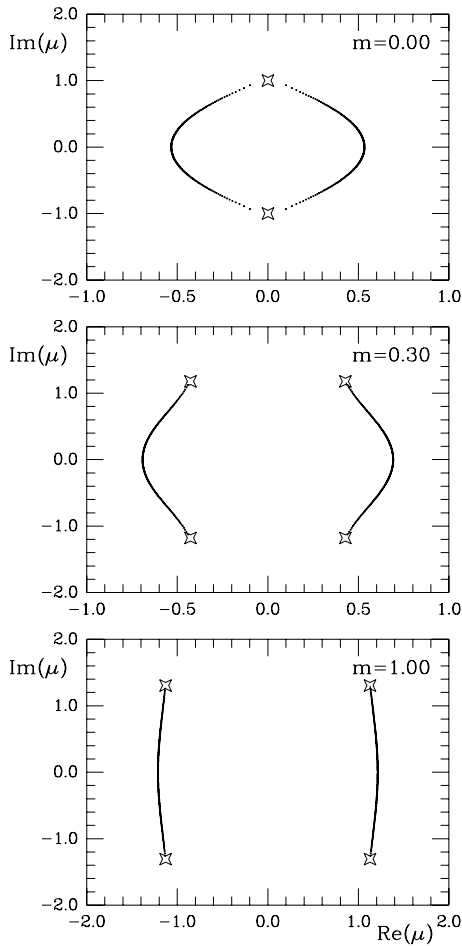,width=60mm}
\caption{The zeros of the partition function in the complex $\mu$ plane
 for $m =0$ (upper), $m = 0.30$ (middle), and $m = 1.0$ (lower) for 
$n = 192$. The solid line in the upper figure represents the solution of 
the transcendental equation (\ref{trans}).  The zeros of the discriminant 
of the cubic equation (\ref{cubic}) are denoted by stars.  Note that the 
scale on the $x$-axis of the lower figure is different.}
\label{lymu}
\vspace*{-0.5cm}
\end{figure}

The stars in Fig.\ 1 represent the points at which the discriminant 
(\ref{discriminant}) of the cubic saddle point equation vanishes. These 
points coincide with the endpoints of the line of zeros.  We have verified 
that the line of zeros coincides with the line at which the partition 
function is dominated by a different solution of the third-order equation. 
A schematic picture of this line is shown in \cite{stephanovlat}.
At the critical value of $\mu$, the line of zeros splits into two lines.

In  Fig. 2 we show the zeros of the partition function
in the complex $\mu$ plane for n=192 and masses
$m = 0$, $ m = 0.3$ and $m= 1.0$.
We have evaluated the zeros for $n=48$ and $n=96$ as well. Also in this case
we have performed our calculations with 100-500 digits accuracy. The zeros
are regularly spaced and their density increases homogeneously with $n$. In
the thermodynamic limit, we therefore expect that they join into a cut. In all
cases the endpoint of the line of zeros ends in a zero of the discriminant
(\ref{discriminant}) 
(denoted by a star). For $m =0$ the thermodynamic limit of the
line of zeros is given by the solution of the transcendental equation
(\ref{trans}) (full line in upper figure).
Chiral symmetry is broken in the region
enclosed by this curve and is restored outside.
Up to corrections of order $1/n$ the zeros coincide with this curve.
For $m = 0$ the density of zeros approaches zero near $\mu = i$.  This is not
surprising since at this point the phase transition changes from  first order
into second order.

4.  In conclusion, we have evaluated the zeros of the partition
function in a schematic random matrix model of the chiral transition
at non-zero chemical potential.  Since the coefficients of these
polynomials were obtained analytically, it was possible to locate
the zeros numerically with considerable accuracy provided that all
arithmetic operations were executed with a precision of 100 to 500
digits.  (We stress that calculations with ordinary double or quadruple
precision arithmetic are not adequate.)  We have found that these zeros lie
on one-dimensional subsets in the complex $m$ or $\mu$ planes.  All
of our results are consistent with a mean field analysis of the random
matrix partition function.  In particular, we have obtained an analytic
expression for the curve of zeros in the complex $\mu$ plane.

{\bf Acknowledgements}
\vglue 0.4cm
This work was partially supported by the US DOE grant
DE-FG-88ER40388. We thank Robert Shrock for useful discussions and educating us
on to general properties of Yang-Lee zeros. James Osborn is thanked for 
pointing
out the existence of multiprecision packages. Melih Sener is thanked for useful
discussions. Finally, we acknowledge
D.H. Bailey and NASA Ames for making their  multiprecision package available.

\setlength{\baselineskip}{14pt}

\newpage
\setlength{\baselineskip}{21pt}
\noindent
{\large\bf Figure Captions}
\vspace*{ 1cm}

\noindent
Fig.1. The zeros of the partition function in the complex $m$ plane
for $\mu =0$ (upper), $\mu = 0.50$ (middle) and $\mu = 0.60$ (lower) 
for $n = 192$.  The zeros of the discriminant of the cubic equation 
(\ref{cubic}) are denoted by stars.

\vspace*{ 1cm}

\noindent
Fig.2 {The zeros of the partition function in the complex $\mu$ plane
 for $m =0$ (upper), $m = 0.30$ (middle), and $m = 1.0$ (lower) for 
$n = 192$. The solid line in the upper figure represents the solution of 
the transcendental equation (\ref{trans}).  The zeros of the discriminant 
of the cubic equation (\ref{cubic}) are denoted by stars.  Note that the 
scale on the $x$-axis of the lower figure is different.}

\end{document}